\documentclass[sn-mathphys,Numbered,pdflatex]{sn-jnl}

\usepackage{graphicx}%
\usepackage{multirow}%
\usepackage{amsmath,amssymb,amsfonts}%
\usepackage{amsthm}%
\usepackage{amssymb}
\usepackage{mathrsfs}%
\usepackage[title]{appendix}%
\usepackage{xcolor}%
\usepackage{textcomp}%
\usepackage{manyfoot}%
\usepackage{booktabs}%
\usepackage{algorithm}%
\usepackage{algorithmicx}%
\usepackage{algpseudocode}%
\usepackage{listings}%
\usepackage{upgreek}

\usepackage{etoolbox}   
\makeatletter
\patchcmd{\@maketitle}{\artauthors}{{\artauthors}}{}{}
\makeatother

\begin{document}

\title[Article Title]{Modelling turbulent flow of superfluid $^{4}$He past a rough solid wall in the $T =$ 0 limit}


\author*[1]{\fnm{Matthew J.} \sur{Doyle}}\email{matthew.doyle@manchester.ac.uk}

\author[1]{\fnm{Andrei I.} \sur{Golov}}\email{andrei.golov@manchester.ac.uk}

\author[1]{\fnm{Paul M.} \sur{Walmsley}}\email{paul.walmsley@manchester.ac.uk}

\author[2]{\fnm{Andrew W.} \sur{Baggaley}}\email{andrew.baggaley@newcastle.ac.uk}

\affil[1]{\orgdiv{Department of Physics and Astronomy}, \orgname{University of Manchester}, \orgaddress{\street{Schuster Building}, 
\city{Manchester}, \postcode{M13 9PL}, \country{United Kingdom}}}

\affil[2]{\orgdiv{Joint Quantum Centre (JQC) Durham–Newcastle}, \orgname{School of Mathematics and Statistics}, \orgaddress{\street{Newcastle University}, \city{Newcastle upon Tyne}, \postcode{NE1 7RU}, \country{United Kingdom}}}


\abstract{We present a numerical study, using the vortex filament model, of vortex tangles in a flow of pure superfluid $^4$He in the $T = 0$ limit through a channel of width $D = 1$\,mm for various applied velocities $V$. The flat channel walls are assumed to be microscopically rough such that vortices terminating at the walls are permanently pinned; vortices are liberated from their pinned ends exclusively through self-reconnection with their images. Sustained tangles were observed, for a period of 80\,s, above the critical velocity $V_c \sim 0.20$\,cm\,s$^{-1} = 20 \kappa/D$. The coarse-grained velocity profile was akin to a classical parabolic profile of the laminar Poiseuille flow, albeit with a non-zero slip velocity $\sim$ 0.20\,cm\,s$^{-1}$ at the walls. The friction force was found to be proportional to the applied velocity. The effective kinematic viscosity was $\sim 0.1\kappa$, and effective Reynolds numbers within $\mathrm{Re’} < 15$. The fraction of the polarized vortex length varied between zero in the middle of the channel and $\sim$ 60\% within the shear flow regions $\sim D/4$ from the walls. Therefore, we studied a state of polarized ultraquantum (Vinen) turbulence fuelled at short lengthscales by vortex reconnections, including those with vortex images due to the relative motion between the vortex tangle and the pinning rough surface.}




\maketitle

\section{Introduction}\label{sec1}

Flow of superfluid $^4$He through a channel is only non-dissipative when its velocity is below the channel-specific critical velocity $v_c$. Above it, the dissipative regime sets in, in which the chaotic motion of quantum vortices (called Quantum Turbulence) results in the transfer of the flow momentum to the channel's walls, i.\,e. a friction force. 

    Experimental evidence has shown that this friction force becomes greatly reduced for $T<$ 0.7 K \cite{Zmeev2015} and that vortex pinning becomes much weaker with lowering temperature below $T < 0.4$ K \cite{Donev2001}. 
        Approaching $T = 0$, the density of viscous normal component vanishes and the mutual friction which couples it to the vortex tangle of turbulent superfluid is negligible, thus the direct interaction between vortex lines and irregularities of the walls of the channel  (vortex pinning) must be considered. 
        
    Early evidence of quantized vorticity sticking to rough surfaces was observed in 1958 by Hall and Shoenberg in torsional-oscillator experiments \cite{Hall1958}. Further studies involving rotating quantum turbulence \cite{Adams1985-Spin-upProblem}, thermal counterflow \cite{Hedge1980} and vortex capture probes such as wires \cite{Neumann2014} and MEMS devices \cite{Barquist2020} aid in providing confirmation of the influence of surface roughness and vortex pinning on superfluid flows. Numerical studies have examined vortex motion in the presence of solid boundaries through use of the vortex filament model (VFM) \cite{Schwarz1981, Schwarz1985, Schwarz1988, Schwarz1992, Schwarz1993, Tsubota1993, Tsubota1994, Fujiyama2009, Nakagawa2023} and the Gross-Pitaevskii model \cite{Stagg2017SuperfluidBoundaryLayer}.

In these computer simulations, we model a flow of superfluid $^4$He at $T=0$ between two parallel solid walls. We assume the limit of extremely rough walls, where the areal density of sharp protuberances is greater than the density of vortex lines in the vortex tangle. Then the processes of un-pinning (due to a self-reconnection with an image vortex) of each vortex line from the pinning protuberance and re-pinning at the nearest protuberance occur on the lengthscale smaller than the characteristic scale of the vortex tangle (of order mean inter-vortex distance $\ell$), i.\,e. independently of other vortices. This is different from the the generation of dense vortex tangles by larger-scale irregularities of the profile of solid wall simulated by Stagg et al. \cite{Stagg2017SuperfluidBoundaryLayer}.

In numerical simulations vortex lines are represented by chains of discrete points with the set inter-point scale $\delta$, the smallest resolved lengthscale is $\sim \delta$. Within the mechanism of pinning-unpinning mentioned above, the elementary distance between the unpinned-pinned ends of a vortex line becomes of order $\sim \delta$. Hence, in our framework, $\delta$ is a proxy of the scale of roughness of the solid wall.
    

\section{Numerical Methods}\label{sec2}
\begin{figure}[h]%
        \centering
        \includegraphics[trim={1cm 1cm 1cm 1cm},width=0.99\textwidth]{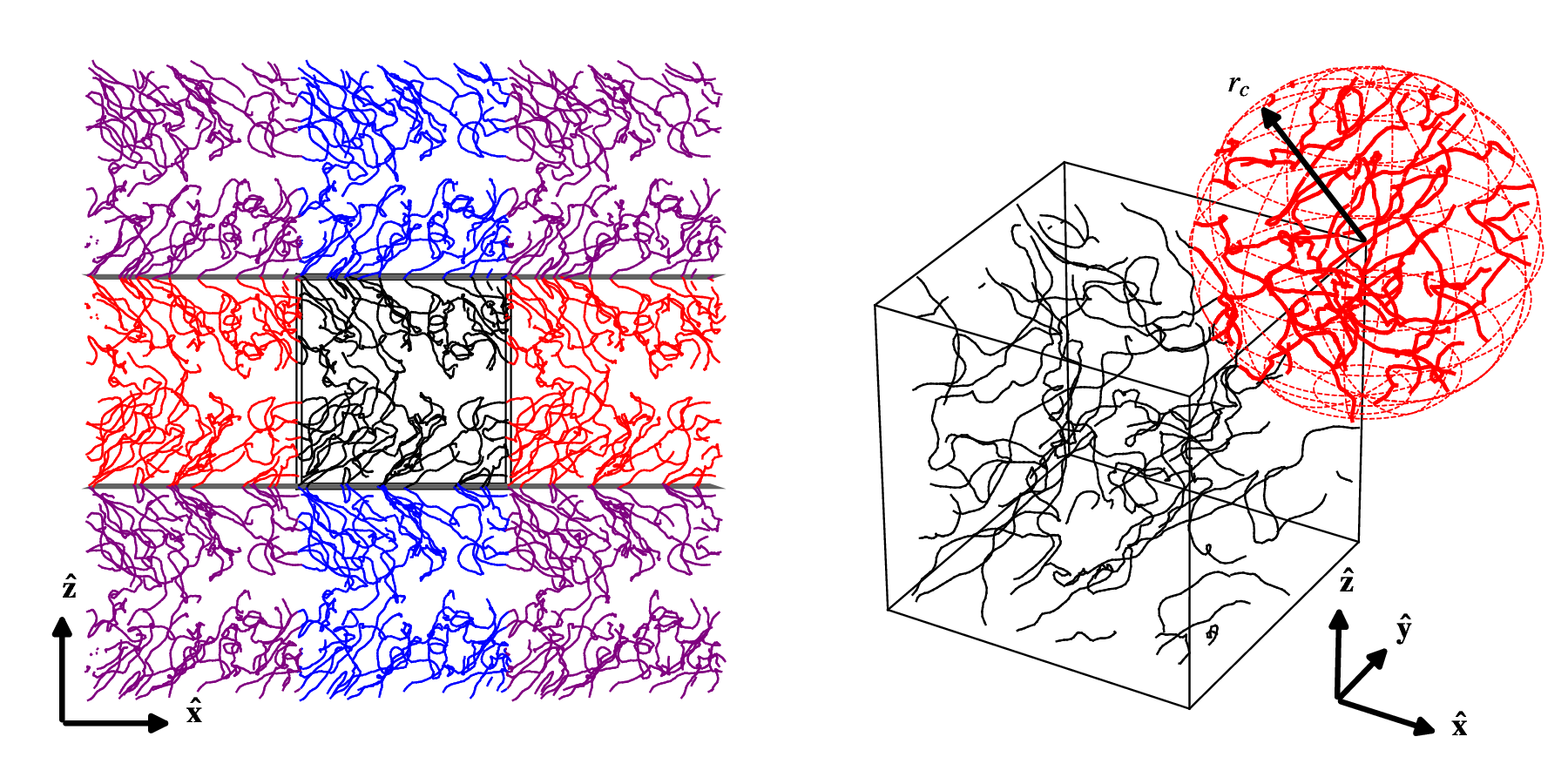}
        \caption{(Left) Visualisation of simulated boundary conditions. The periodicity of the boundaries is shown only in the $x$ direction such that 9 out of 27 cubes are displayed. The original computational volume in the centre is shown by the black box outline. Flat grey surfaces represent the solid boundaries and are extended into the periodic boundaries in the $x$-direction. The vortex tangle in each cube is coloured to indicate the type of transformation applied: periodic-translation (red), solid-refelction (blue) and a combination of the two (purple). (Right) Illustration of the sphere of velocity contributions centred on a point, $\mathbf{s}_i$, near the corner of the volume. Red vortex segments, within the sphere, contribute to $\mathbf{\Dot{s}}_i$ and vortex lines outside of the sphere are neglected. }
        \label{fig1}
    \end{figure}

    The VFM \cite{Hanninen2014VFM} has been used to great effect to simulate and visualise the dynamics of vortices in superfluid helium. 
    In the limit of zero temperature, using the VFM, the local self-induced vortex velocity $\Dot{\mathbf{s}}$ evaluated at the point $\mathbf{r}$ along a vortex line may be described entirely by the Biot-Savart integral
    \begin{equation}
        \Dot{\mathbf{s}} (\mathbf{r}, t) = \frac{\kappa}{4\pi} \int_\mathcal{L} \frac{ ( \mathbf{s} - \mathbf{r} ) \times d\mathbf{s} }{ | \mathbf{s} - \mathbf{r} |^3},
    \end{equation}
    where $\kappa = h / m_4$ is the quantum of circulation in superfluid $^4$He. The line integral over $\mathcal{L}$ represents inclusion of the complete vortex configuration, which is discretized into points $\mathbf{s}_i$ for $(i = 1,...,N)$. The discretized integral, after removing the singularity by separating the local and non-local contributions, becomes 
    
    \begin{equation}
       \Dot{\mathbf{s}}_i = \frac{\kappa}{4\pi} \mathbf{s}_i' \times \mathbf{s}_i'' \text{ln} \bigg( \frac{2\sqrt{l_+l_-}}{e^{1/2}a} \bigg) + \frac{\kappa}{4\pi} \int_{r_c} \frac{(\mathbf{s}_1 - \mathbf{s}_i) \times d\mathbf{s}_1}{|\mathbf{s}_1 - \mathbf{s}_i|^3},
       \label{eq:2}
    \end{equation}
    where $l_+$ and $l_-$ represent the arc lengths to adjacent vortex points $\mathbf{s}_{i+1}$, and $\mathbf{s}_{i-1}$, $\mathbf{s}_i'$ and $\mathbf{s}_i''$ are the local tangent and curvature respectively, and $a \sim$ 1 \AA \space is the size of the vortex core in He-II. Here the first term is akin to the local induction approximation (LIA) and the second term describes the non-local contributions, including the effect of image-vortices due to any solid boundaries present \cite{Schwarz1985}. To reduce the number of computations required to simulate vortex tangles in a timely manner, the integral has been reduced to only include contributions from points within a cut-off radius $r_c$ of the target point $\mathbf{s}_i$, defining a sphere of contributing non-local vortex points as shown in Figure \ref{fig1}\,(right). The cut radius was chosen to be half the container size $r_c = D/2$. 
    

    %

    The simulation volume was a cubic cell $D\times D \times D$ with periodic boundary conditions in the $x,y$ directions. The boundaries at $z=0$ and $z=D$
    were modelled as a rough solid plane surface with strong pinning. To satisfy the solid boundaries the method of images allows the duplication and reflection of the vortex configuration across each solid boundary. This combined with a periodic wrapping of the system at the $x,y$ boundaries gives 26 copies of the original volume: 2 reflected at solid boundaries, 8 simply periodic and 16 periodic-reflected copies. An illustration of the boundary conditions in the $x$ and $z$ directions is given in Fig.\,\ref{fig1}\,(left). All 27 cubes were considered in evaluating Equation \ref{eq:2}.

    The spatial resolution $\delta$ of the simulation is preserved by removing points on the same vortex line that move within $\delta/2$ of each other. Similarly, new points are introduced, according to the local curvature of the line, when adjacent points are more than $\delta$ apart. Additionally, if two non-adjacent points on the same vortex line or otherwise, become closer than the critical distance $\delta/2$, a line-line reconnection occurs, altering the vortex configuration and causing dissipation from the removal of vortex line length. The reconnection procedure used was of Type-III \cite{Baggaley2012sensitivity}. 

    To produce channel flow in the $T=0$ limit a positive finite superfluid velocity $V$ is applied in the $x$ direction such that instantaneous velocity of the discrete vortex point is 
    \begin{equation}
        \mathbf{u}_i = \Dot{\mathbf{s}}_i + 
            \begin{bmatrix}
               V \\
               0 \\
               0
            \end{bmatrix}.
    \end{equation}

    The time-evolution followed a third order Adams-Bashforth scheme:
    \begin{equation}
        \mathbf{s}_i^{n+1} = \mathbf{s}_i^n + 
        \frac{\Delta t}{12} \bigg( 23 \mathbf{u}_i^n - 16 \mathbf{u}_i^{n-1} + 5 \mathbf{u}_i^{n-2} \bigg) + \mathcal{O}(\Delta t^4)
    \end{equation}
   where $\Delta t$ is the size of a time step and $n = t / \Delta t$ is the current time in integer steps \cite{TreeMethod}. 

\begin{figure}[h]%
        \centering
        \includegraphics[trim={1cm 1.5cm 1cm 1.5cm},width=0.6\textwidth]{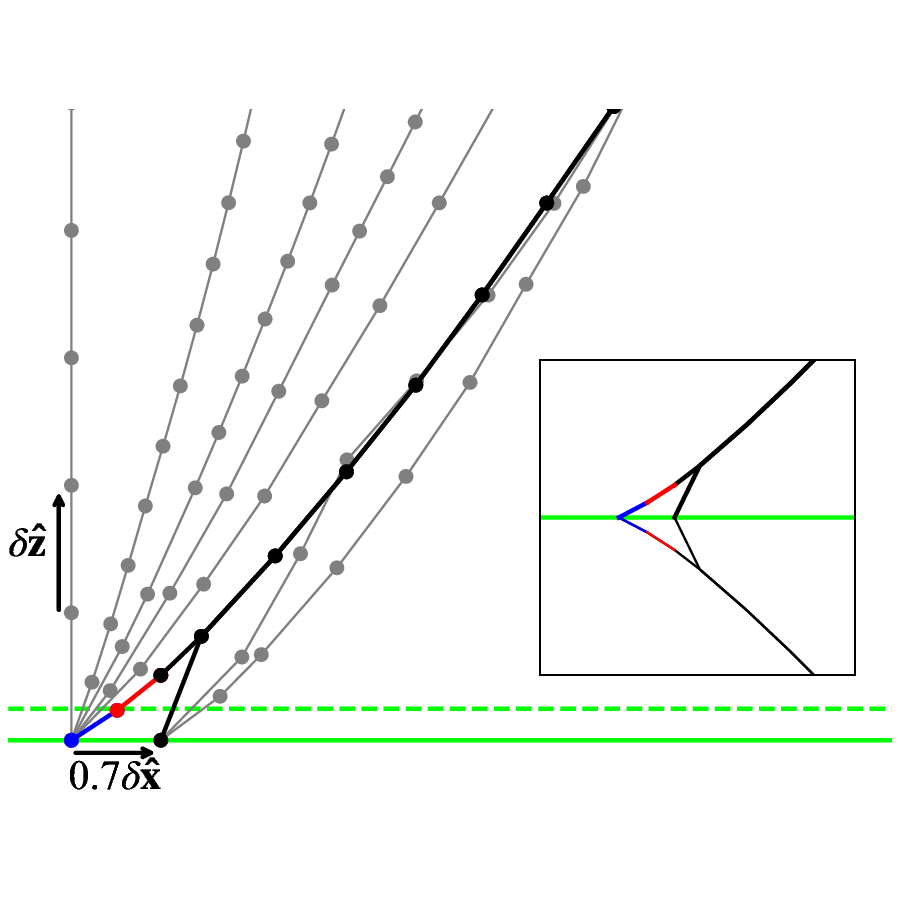}
        \caption{Progressive snapshots of a single vortex line, which is initially pinned vertically between two flat rough surfaces, evolving under an applied velocity $V$ = 0.05 cm s$^{-1}$. Vortex lines drawn in grey are shown at times: 50\,ms, 100\,ms, 150\,ms, 200\,ms, 255\,ms and 260\,ms. The region displayed is the lower 5\% of the channel with the solid green line marking the solid boundary. The dashed green line shows the critical distance for vortex-wall reconnection. The solid black lines show the filament in the instant before and after the first time the vortex reconnected with the wall at $t=253.69$\,ms. Coloured line segments (red and blue) show the length removed by the event, and the blue segment is responsible for the reconnection. The size of arrows indicates the spatial resolution, $\delta$ and the distance between new and old pinning sites. The inset illustrates the moment of reconnection, with the image-vortex drawn symmetrically behind the boundary.
        }
        \label{fig2}
    \end{figure}

   The presence of solid wall boundaries alters the vortex motion since no flow is allowed through the walls. The solid boundaries are assumed to be microscopically rough such that the ends of vortex lines, terminated at the walls, are permanently pinned and fixed in position with $\mathbf{u}(z=\pm D/2)$ = 0. 
   Liberation of the vortex line is only possible through its self-reconnection with its image following the same reconnection criterion that triggers a line-line event. Thus, a vortex point, which comes within a distance $\delta$⁄4 of the wall, reconnects with its image, thus liberating the line before it becomes pinned again $\sim 0.7\delta$ away, as demonstrated in Figure \ref{fig2}. Here, the reconnection was triggered by the blue line segment, which is then removed from the simulation alongside the adjacent red line segment. The new end of the vortex is then instantaneously pinned by artificially adjusting the $z$ coordinate to coincide with the surface, which generates a kink in the filament just above the wall. The model mimics a vortex line “walking” along a flat rough surface, with the vortex end jumping between sharp protuberances spaced on the scale of the spatial resolution $\delta$ \cite{Golov2021qfs}. 

Effectively, this mechanism assumes the existence of a critical angle between the vortex and the wall for unpinning. 
Such an assumption was recently used in simulations of vortex lines attached to a microelectromechanical (MEMS) oscillator, with a chosen critical angle of $\theta_c = \pi/6$ \cite{Nakagawa2023}.

The potential benefits of our approach are in that one could investigate the role of Kelvin waves emitted on vortex lines by discrete steps; although to assure an adequate resolution of those waves one might need to introduce a separate lengthscale for the kink generated by an image-reconnection such that the wavelength of injected distortions is larger than the resolution. 

    To examine the rate of momentum transfer to the walls, two equivalent methods of determining the friction force were used. Both have been validated by comparison with the analytical solution for a vortex semi-ring travelling along a flat rough surface. 
    The rate of change of the total impulse of all vortex lines in the simulation volume, termed the {\it integral} method: 
    \begin{equation}
        \mathbf{F} = \rho_s \kappa \int (\Dot{\mathbf{s}} \times \mathbf{s}' ) d\xi,
        \label{eq-integral}
    \end{equation}
    where $\rho_s$ is the superfluid density and $\xi$ is the arc length along a vortex line. 
    The second method assumes that the vortex line tension $f_t$ (energy per unit line length) is constant. The angle $\theta$ between the surface and the pinned segment can be used to obtain the components of the line tension for a single terminated line. Thus, summing over all the pinned ends gives the total friction force in the streamwise direction
    \begin{equation}
        F_x = -f_t \sum_{i,pinned} \text{cos} \theta_i = - \frac{\rho \kappa^2}{4\pi} \text{ln} \Big( \frac{b}{a} \Big) \sum_{i,pinned} \text{cos} \theta_i ,
        \label{eq-tension}
    \end{equation}

    where $b$ was a cut-off length scale chosen as $\delta/2$ which is the effective size of pinning sites in the vortex walking  and $a = 1$\,\AA. We term this the {\it tension} method.  

    To examine the velocity profile, the instantaneous mean velocity as a function of the wall-normal direction $z$ must be calculated.
    This was achieved by slicing the simulation volume into $N$ slabs with thickness $D/(N-1)$, the mean velocity of all line segments within a single slab was then time-averaged during the steady state to obtain the coarse-grained velocity profile $\langle \mathbf{u}(z) \rangle$. 
    The component in the direction of flow, $\langle u_x \rangle$ then gives the cross-channel velocity profile. 
    Fixed segments of the vortex line were omitted from such calculations to avoid skewing the average velocity, with $u_{pinned}=0$, in the top and bottom slabs.  
    The variation in the coarse-grained vortex line density across the height of the channel $\langle \mathcal{L}(z) \rangle$ was obtained similarly, but without the need to exclude fixed segments.

\section{Results}\label{sec5}
    
\begin{figure}[h]%
        \centering
        \includegraphics[trim={2cm 1cm 2cm 1cm},width=1\textwidth]{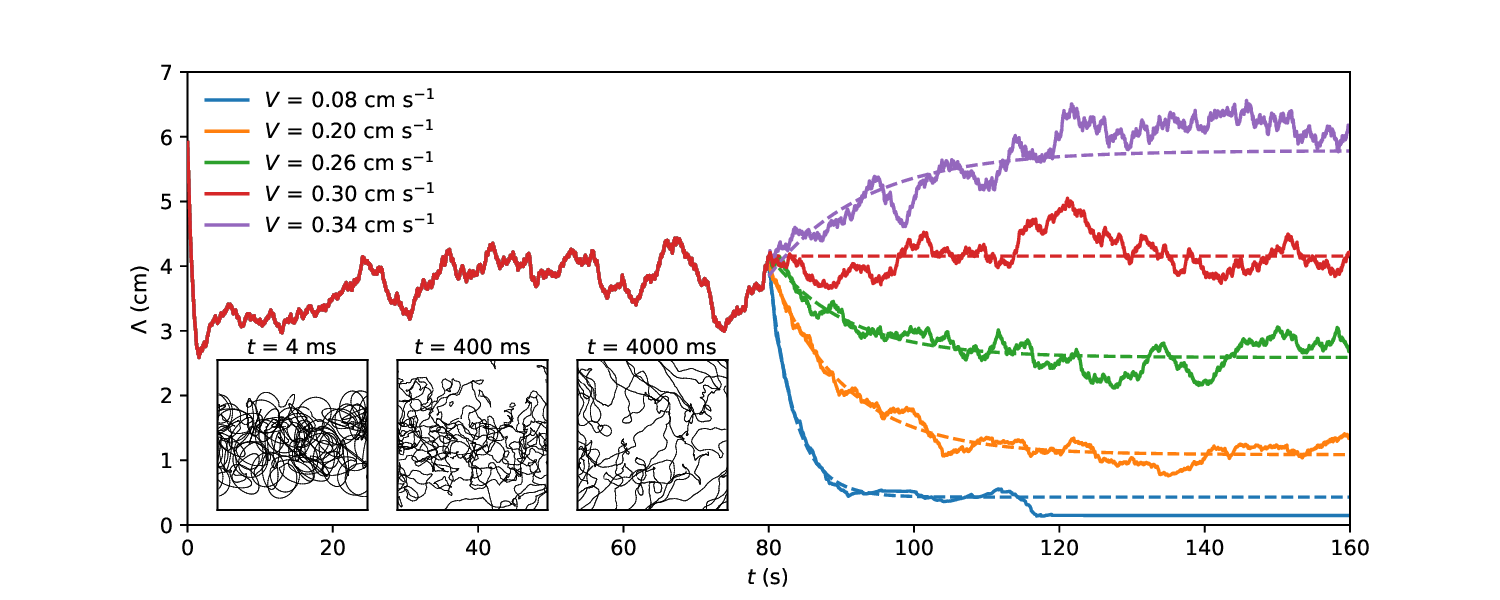}
        \caption{Vortex line length as a function of time shown for 5 different simulations. Solid lines show data and dashed lines show fitted curves. The inset plots are early-time snapshots of the developing vortex tangle for $V=0.30$\,cm\,s$^{-1}$.}\label{fig3}
    \end{figure}

    The parameters for simulations were: $D = 0.1$\,cm, $\delta = 2\times 10^{-3}$\,cm, with temporal resolution $dt = 8\times 10^{-5}$\,s. 
    Initially, at $t=0$, the volume was populated with 80 randomly placed and randomly oriented vortex rings with equivalent radii of 0.012\,cm; with an instantly acting applied superflow of $V = 0.30$\,cm s$^{-1}$. 
    Snapshots of the early development of the tangle is shown by the inset panels of Fig.\,\ref{fig3}. 
    At early times ($t<1$\,s) frequent reconnections between vortex rings in the centre of the channel cause an immediate drop in the total vortex line length, $\Lambda$, of the system, which is also shown in Fig. \ref{fig3}. 
    Gradually vortex rings become attached to the walls and begin ``walking'' across the surface in the direction of flow. Within $t = 4$\,s both surfaces are decorated with several terminating vortices. 

    As the flow develops further vortex lines become stretched along it causing a gradual increase in the total line length as a vortex tangle forms between the walls. 
    Both line-line and image-line reconnections occur frequently, each process removing a small portion of vortex length, which becomes approximately stable when the rate of dissipating length from reconnections is similar to the rate of vortex growth at the walls.
    At $t = 80$\,s the vortex tangle is comfortably in the steady state with the total vortex length in the simulation volume $\Lambda = 4.14$\,cm, and the walls are decorated with 14 pinned vortex lines. The tangle at $t = 80$\,s is the same as the configuration shown in Fig. \ref{fig1}.

    To simulate flows at any other applied velocity in the range 0.08\,cm\,s$^{-1} \leq V \leq$\,0.34\,cm\,s$^{-1}$ for further 80\,s, the same initial configuration of the vortex tangle was used, namely, the one for $V=0.30$\,cm\,s$^{-1}$ at $t=80$\,s. 
 
\begin{figure}[h]%
        \centering
        \includegraphics[width=0.99\textwidth]{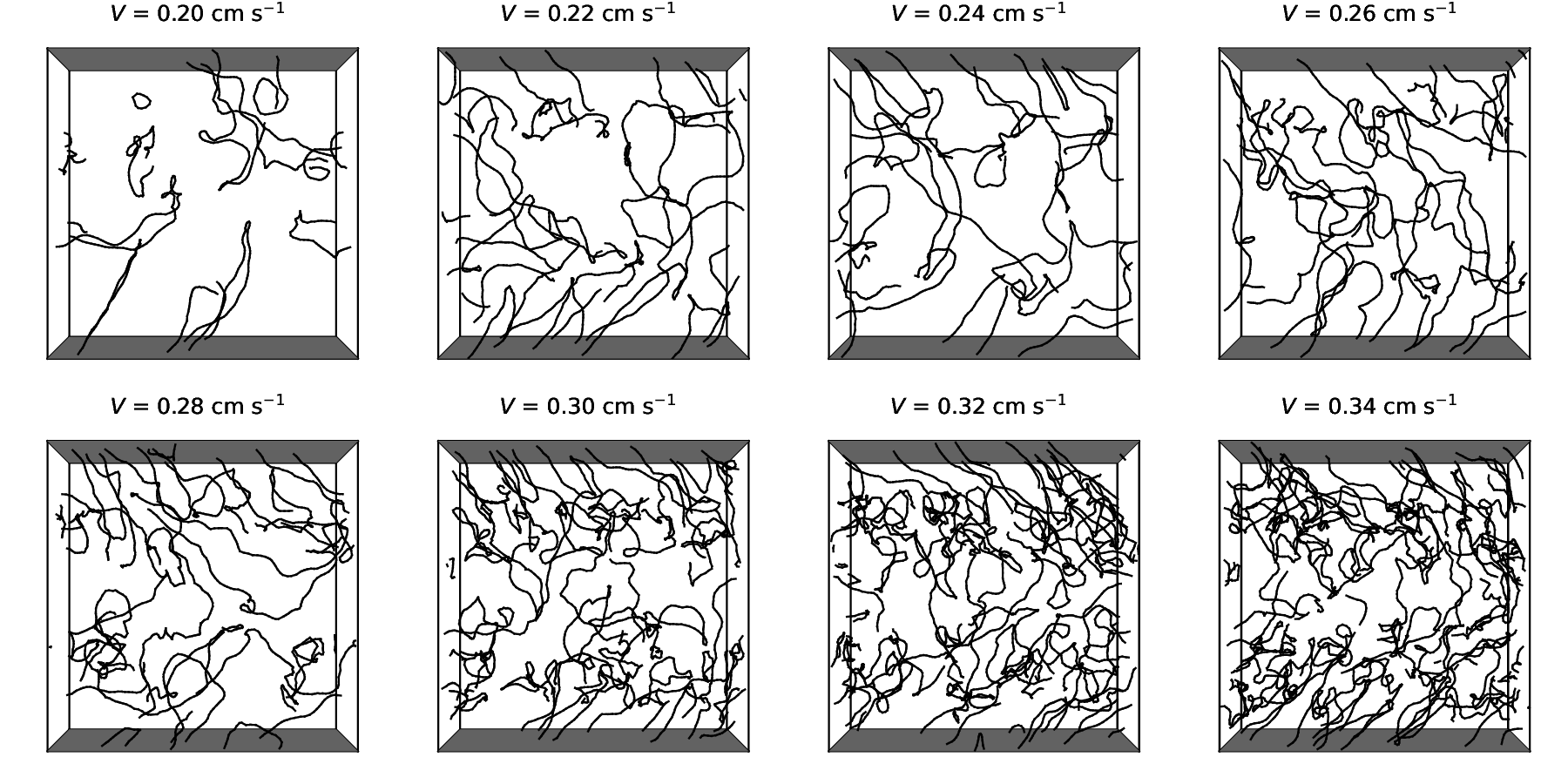}
        \caption{Snapshots of vortex tangles in the steady state at $t=$ 160 s for all $V > V_c$. The flow is directed towards the right.}\label{fig4}
    \end{figure}

    The evolution of $\Lambda$ for a selection of $V$ is shown in Fig.\,\ref{fig3} in the range 80\,s\,$\leq t \leq$\,160\,s.
    Vortex tangles, sustained for at least 80\,s as in Fig. \ref{fig4}, were observed above the critical applied flow velocity $V_c = 0.19 \pm 0.01$ cm\,s$^{-1}$, i.\,e. $V_c D/\kappa \sim 20$. Below $V_c$, the tangle is not sustainable; all vortex lines eventually detach from the walls at some point and move with the flow without gaining any energy -- hence the tangle decays.

\begin{figure}[h]%
        \centering
        \includegraphics[width=.99\textwidth]{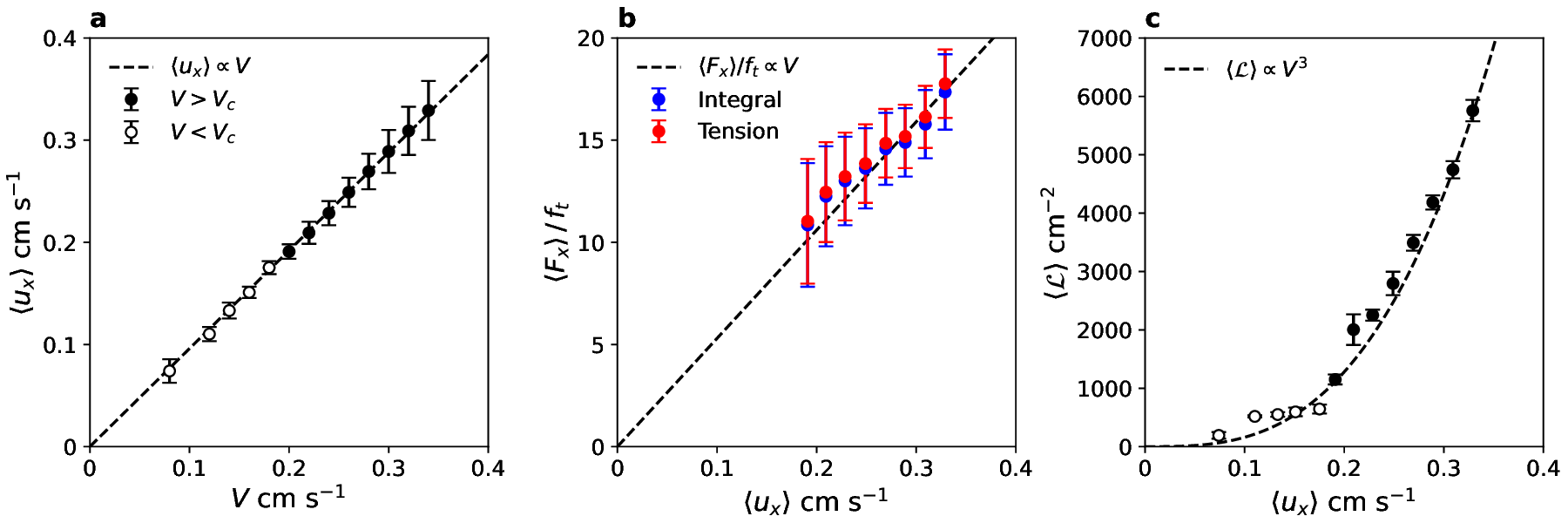}
        \caption{(a) Time-averaged mean flow velocity as a function of the applied superflow shown for both sustained (closed symbols) and unsustained (open symbols) turbulence. (b) The coarse-grained friction force exerted by vortex lines on the channel walls, normalised by a constant vortex line tension, shown to be proportional to the mean flow. (c) Time-averaged vortex line density as a function of the mean flow. Open circles represent unsustainable line densities.
        }
        \label{fig5}
    \end{figure}

    For each applied superfluid velocity $V$, the mean flow velocity was calculated as $\langle u_x \rangle = D^{-1}\int_0^D u_x(z)dz$ and is plotted against $V$ in Fig. \ref{fig5}a. They are proportional to each other with relation $\langle u_x \rangle \approx 0.96 V$.

    The stream-wise component of friction force was evaluated from equations (\ref{eq-integral}) and (\ref{eq-tension}), time-averaged over the steady state and normalised by the reduced vortex line tension $f_t = \frac{\rho \kappa^2}{4\pi} \text{ln} \Big( \frac{b}{a} \Big)$, where $b=\delta /2$ is the effective size of a pinning site. 
    The observed force-velocity relation, shown by Fig. \ref{fig5}b, was approximately linear within the range of flow velocities selected. 
    Friction force per number of pinned vortex ends was constant across all velocities at $(0.626 \pm 0.007)f_t$ demonstrating that the total friction force is proportional to the number of pinned ends in the simulation box and the effective critical angle in the direction of flow is $\gtrsim \cos^{-1}(0.626) = 51^{\circ}$.
    Both {\it integral} and {\it tension} methods of determining the friction force consistently agreed to within 1\%.

The total vortex line length $\Lambda$  fluctuates with the dominant period $\sim 10$\,s (corresponding to the time constant $\sim 2$\,s). This could be compared with the time $D/V_c \sim 0.5$\,s for the mean flow to pass the cell size, the intrinsic quantum time constant $D^2/\kappa = 10$\,s and our chosen observation time of 80\,s.   
    Fig. \ref{fig5}c demonstrates the variation of time-averaged vortex line density, $\langle \mathcal{L} \rangle = \langle \Lambda \rangle / D^3$, which appears to be cubic with velocity.
    Interestingly, the values of $\langle \mathcal{L} \rangle$ for unsustained tangles do not fall far from the $V^3$ line. 
    
\begin{figure}[h]%
        \centering
        \includegraphics[width=0.99\textwidth]{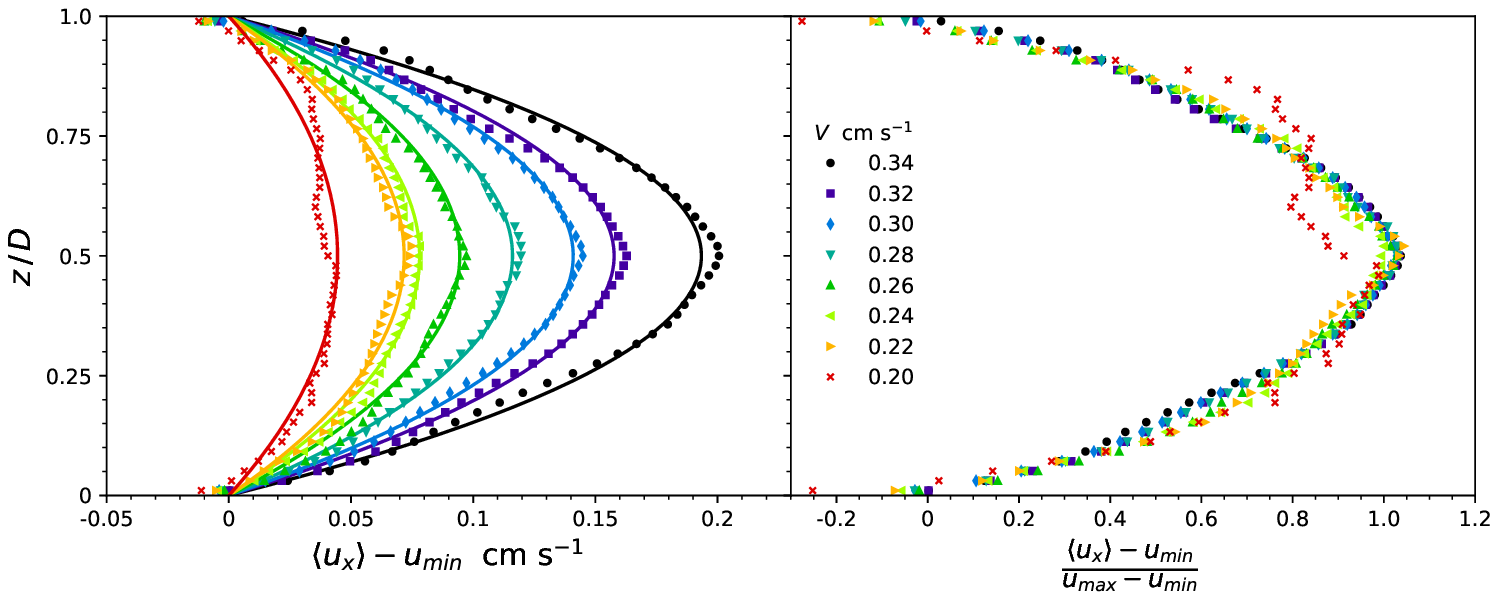}
        \caption{Cross-channel coarse-grained velocity profiles. (Left) Absolute values, reduced by the slip-velocity at the walls. (Right) Normalised by the range of velocity present in the flow.}\label{fig6}
    \end{figure}

\begin{figure}[h]%
        \centering
        \includegraphics[width=0.99\textwidth]{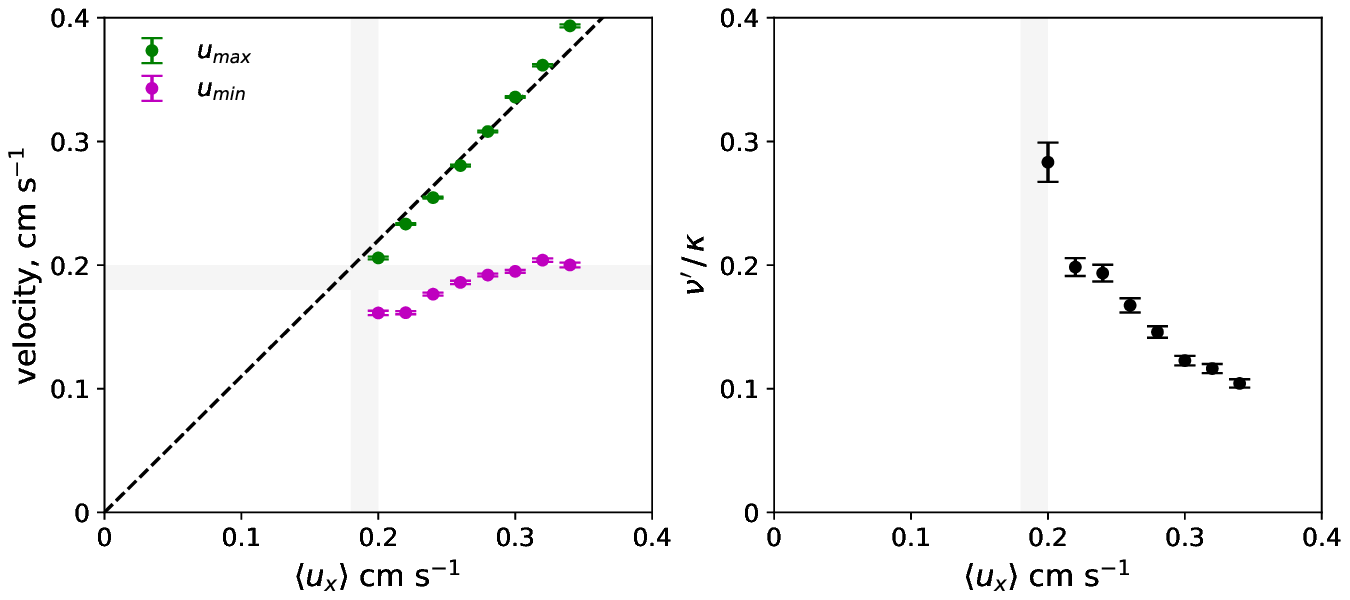}
        \caption{(Left) Results of fitting velocity profiles with equation (\ref{eq-profile}) as function of $\langle u_x \rangle$. Dashed line follows $u_{max} \propto \langle u_x \rangle$. (Right) The effective kinematic viscosity as function of $\langle u_x \rangle$. The vicinity of $V_c$ is shown by the shaded area in both graphs. }\label{fig7}
    \end{figure}

    The coarse-grained stream-wise velocity profile $\langle u_x (z) \rangle$  was steady and nearly parabolic albeit with a non-zero slip velocity, as seen in Fig \ref{fig6}. We fit the profile with the form of a classical Poiseuille profile, adapted for finite slip boundary conditions at the walls, 
    \begin{equation}
        \langle u_x (z) \rangle = u_{max} - 4(u_{max} - u_{min})\left(\frac{z}{D}-\frac{1}{2}\right)^2 \text{,}
        \label{eq-profile}
    \end{equation}
    where $u_{min}$ represents the slip-velocity at the walls and $u_{max}$ is the maximum local velocity in the centre of the channel. The agreement is relatively good, but the fits consistently underestimate $u_{max}$. 
    The slip-velocity varied only weakly across the simulations with values in the region of $V_c$ as seen in Fig.\,\ref{fig7}\,(left). The universality of the flow profiles is shown in Fig.\,\ref{fig6}\,(right). Deviations from the universal profile were only observed for the lowest $V$ which still produced a steady flow - possibly due to the smaller number of line segments leading to less averaging off of fluctuations.     

    The effective kinematic viscosity $\nu' = \frac{\langle F \rangle}{8D(u_{max}-u_{min})\rho}$ was in the range 0.3$\kappa$ – 0.1$\kappa$ -- shown in Fig.\,\ref{fig7}\,(right) -- meaning the effective Reynolds number $\text{Re}' = D(V-V_c)/\nu'$ was between 0 and $\sim 15$. Clearly this was insufficient to support quasi-classical turbulence in the coarse-grained velocity field. Thus, we have polarised ultra-quantum turbulence driven by injections of short-wavelength Kelvin waves, enabled by the frequent reconnection of vortex ends due to the relative motion between the vortex tangle and the rough wall.

\begin{figure}[h]%
        \centering
        \includegraphics[width=0.99\textwidth]{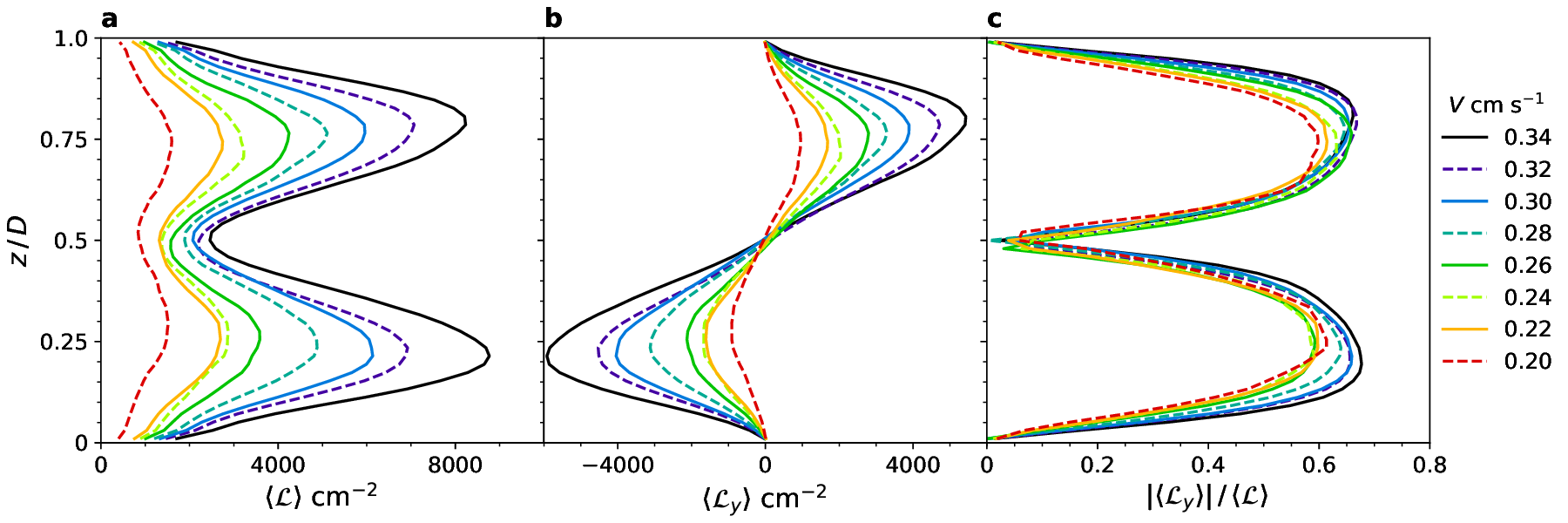}
        \caption{(a) Coarse-grained vortex line density profile. (b) Polarised vortex line density profile. (c) Time-averaged polarisation of the vortex tangle.}\label{fig8}
    \end{figure}

    The coarse-grained vortex line density profile, shown in Fig. \ref{fig8}a, demonstrates the tangle formed maximally dense regions $\sim D/4$ away from each surface, with minima in the centre of the channel and at the walls. Such a spatial variation in $\langle \mathcal{L} \rangle$ is similar with simulations at finite temperature where flows were driven by a normal fluid flow described by a Poiseuille profile \cite{Baggaley2015}. 
    The peaks are mirrored by the polarised line density $\langle \mathcal{L}_y(z) \rangle$ (Fig.\,\ref{fig8}b) and their ratio $\frac{\langle \mathcal{L}_y(z) \rangle}{\langle \mathcal{L} (z) \rangle}$ (Fig.\,\ref{fig8}c) shows the polarisation of the vortex tangle, whose total was between 0 and 60\% depending on the distance from walls, in all  simulations. 
    
\section{Conclusions}\label{sec7}

These simulations demonstrate a critical velocity $V_c = 0.19 \pm 0.01$\,cm\,s$^{-1} \simeq 20\frac{\kappa}{D}$ for pure superfluid flow in channels, above which the turbulent state of the vortex tangle sustained for at least 80\,s. There is a finite slip velocity at the solid boundary, whose value is similar to $V_c$, resultant from the ``walking'' of terminated vortex lines along the rough surface due to frequent reconnections with their images.  The coarse-grained velocity profiles $u_x(z)$ were similar to classical laminar profiles (albeit with a non-zero velocity at the walls) -- consistent with the effective Reynolds number $\text{Re}' \leq 15$ too small to sustain quasi-classical turbulence that typically requires $\text{Re} > 3000$. We thus had ultra-quantum (Vinen's) regime of quantum turbulence. The fraction of the polarised vortex length reached $\sim 60\%$ within the shear regions near both walls, and decreased to zero in the middle of the channel. The effective kinematic viscosity $\nu'$, computed for the shear regions, tends towards $\sim 0.1 \kappa$ with increasing the applied flow velocity -- in quantitative agreement with experimental observations for quantum turbulence in superfluid $^4$He \cite{Zmeev2015}. The flows exhibited an approximately proportional dependence of the friction force $F_x(V)$ and near cubic dependence of the vortex line length $\Lambda(V)$ on the mean flow velocity $V$. These dependences held even below $V_c$ -- for metastable vortex tangles of lifetime shorter than 80\,s.

\bmhead{Acknowledgments}
The authors would like to acknowledge the assistance given by Research IT and the use of the Computational Shared Facility at The University of Manchester. This work was supported by the Engineering and Physical Sciences Research Council [grant number EP-T517823-1]. 

\bmhead{Data Availability}
The code used to generate these simulations and guidance in its use can be accessed from the authors upon reasonable request.


\bibliography{sn-bibliography.bib}

\end{document}